\documentclass[conference]{IEEEtran}
\IEEEoverridecommandlockouts
\usepackage{cite}
\usepackage{amsmath,amssymb,amsfonts}
\usepackage{algorithm}
\usepackage{algorithmicx}
\usepackage{algpseudocode}

\usepackage{graphicx}
\usepackage{textcomp}
\usepackage{xcolor}

\begin{document}

\title{DMCE: Diffusion Model Channel Enhancer for Multi-User Semantic Communication Systems
\thanks{This work was supported in part by the National Natural Science Foundation of China under Grant 62001051 and in part by the China Scholarship Council.}
\vspace{-0.2cm}
}

\author{\IEEEauthorblockN{Youcheng Zeng$^1$, Xinxin He$^1$, Xu Chen$^{1,2}$, Haonan Tong$^{1,3}$, Zhaohui Yang$^4$, Yijun Guo$^1$, Jianjun Hao$^1$ }
\IEEEauthorblockA{$^1$School of Information and
Communications Engineering, Beijing University of Posts and Telecommunications,\\
Beijing, China\\
$^2$China Academy of Information and Telecommunications Technology (CAICT), Beijing, 100191, China\\
$^3$School of Computer Science and Engineering, Nanyang Technological University, Singapore\\
$^4$College of Information Science and Electronic Engineering, Zhejiang University, Hangzhou, China\\
Emails: \{zeng9910, hxx\_9000, chenxu96330, hntong, guoyijun, jjhao\}@bupt.edu.cn, yang\_zhaohui@zju.edu.cn
}
\vspace{-1cm}
}
\maketitle

\begin{abstract}
To achieve continuous massive data transmission with significantly reduced data payload, the users can adopt semantic communication techniques to compress the redundant information by transmitting semantic features instead. However, current works on semantic communication mainly focus on high compression ratio, neglecting the wireless channel effects including dynamic distortion and multi-user interference, which significantly limit the fidelity of semantic communication. To address this, this paper proposes a diffusion model (DM)-based channel enhancer (DMCE) for improving the performance of multi-user semantic communication, with the DM learning the particular data distribution of channel effects on the transmitted semantic features. In the considered system model, multiple users (such as road cameras) transmit semantic features of multi-source data to a receiver by applying the joint source-channel coding (JSCC) techniques, and the receiver fuses the semantic features from multiple users to complete specific tasks. Then, we propose DMCE to enhance the channel state information (CSI) estimation for improving the restoration of the received semantic features. Finally, the fusion results at the receiver are significantly enhanced, demonstrating a robust performance even under low signal-to-noise ratio (SNR) regimes, enabling the generation of effective object segmentation images. Extensive simulation results with a traffic scenario dataset show that the proposed scheme can improve the mean Intersection over Union (mIoU) by more than 25\% at low SNR regimes, compared with the benchmark schemes.

\end{abstract}

\vspace{-0.3cm}
\section{Introduction}
Semantic communication, as an information transmission technique that transmits semantic features instead of coded symbols \cite{chen2023foundation}, is a promising scheme to enable ubiquitous massive amount data transmission in future networks. Despite widespread studies on semantic communication with perfect physical channel estimation and equalization\cite{tong2021}, multi-user systems still lack robust mechanisms to manage dynamic channel noise and multi-user interference in the absence of complete channel information, which is essential for ensuring communication fidelity \cite{lalouani2022federated, Yang2023Energy}.

Extensive works \cite{Zhang2022, huang2021deep6, ZHANG2021, Luo202218, zhang2022Multi-user} have been conducted on semantic communication systems for image transmission. The authors in \cite{Zhang2022} proposed a multi-level semantic-aware wireless image transmission system, encoding text semantics and segmentation semantics at higher-level and lower-level for transmission over wireless channels. The authors in \cite{huang2021deep6} designed a coarse-to-fine image semantic encoding model in which the image semantic features are fully preserved by the base layer, with the enhancement layer restoring the image details. Inspirationally, to overcome the deficiencies of single-source data, the work in \cite{ZHANG2021} effectively fuses multi-source information from the same scene, expressing image semantic information more accurately. The authors in  \cite{Luo202218} proposed a novel multi-user multi-modal semantic communication information fusion scheme, where the receiver can recover semantic information without performing multi-user signal detection. The authors in \cite{zhang2022Multi-user}  introduced a deep learning (DL)-based multi-user semantic communication system for collaborative object recognition, achieving joint recognition by integrating individual semantic features into global features. However, the works in \cite{ZHANG2021, Luo202218, zhang2022Multi-user} have not considered the distortion effects of the dynamic physical channels, and the channel state information (CSI) should be accurately estimated in communications to conduct channel equalization, which compensates for the channel distortion.

 The Denoising Diffusion Probability Model (DDPM) \cite{Ho2020} reconstructs the original image distribution from noisy data by estimating the noise distribution within the image. This concept can be adopted to reduce the impact of channel noise on the accuracy of data recovery. In current research, DDPM has found applications in addressing semantic recovery challenges within wireless communication systems. Furthermore, the authors in \cite{Grassucci2023} transmitted highly compressed semantic information to reduce bandwidth occupation, highlighting that even under noisy channels, diffusion models can generate the images with consistency at the semantic level. The work in \cite{Sengupta2023} introduced a channel sampling method using DDPM, generating samples that are higher fidelity compared to Generative Adversarial Networks (GANs) methods. However, the above works \cite{Grassucci2023, Sengupta2023} mainly regarded DDPM as a generative model in semantic communication systems, while neglecting the effects of channel distortion effects and the channel estimation problem. Which are not suitable for application in multi-user scenarios with interference.
 
\begin{figure*}[t]
\centerline{\includegraphics[width = .73\textwidth]{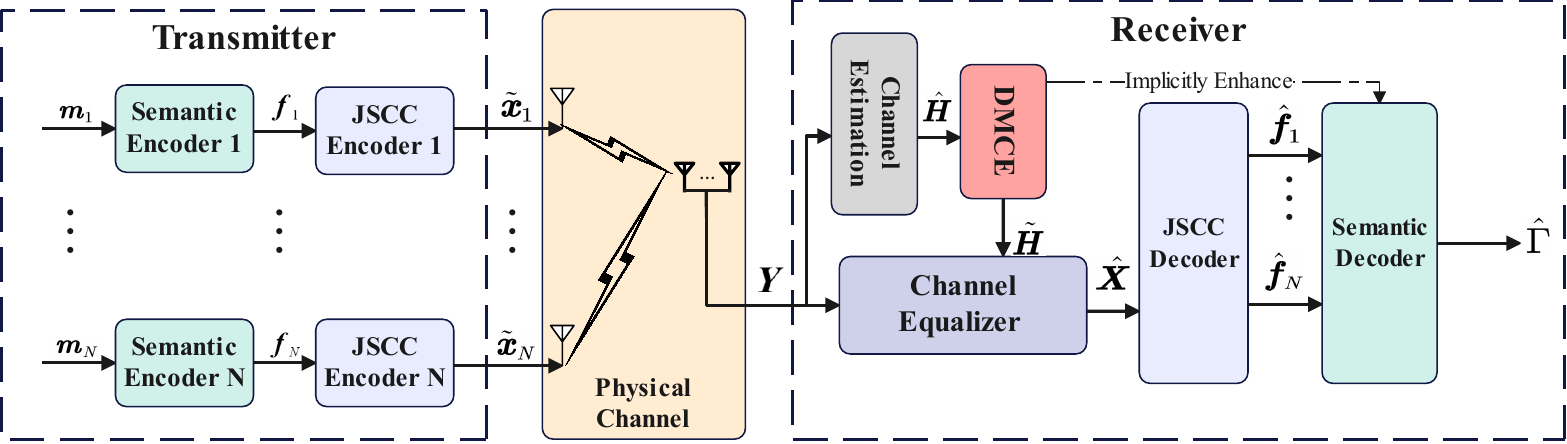}}
\caption{The framework of the proposed DMCE-based multi-user semantic communication systems.}
\label{system structure}
\vspace{-0.6cm}
\end{figure*}

To this end, we propose a DDPM-based channel enhancer at the receiver for the multi-user semantic communication system. This enhances the channel response estimation by suppressing the channel noise, thus enhancing the performance of multisource data fusion via improving the channel equalization at the receiver. To our best knowledge, this is the first work that considers resolving the interference of multi-user signals reception and enhancing the channel estimation to improve the feature fusion performance in the semantic communication system. The primary contributions of this paper are summarized as follows:
\vspace{-0.08cm}
\begin{itemize}
\item We develop a multi-user semantic communication system for the fusion of multi-source traffic scene images, e.g., RGB and IR. Each user needs to transmit the extracted semantic features through a Multiple-Input Multiple-Output (MIMO) channel, and the receiver fuses the multi-source semantic features to recover the corresponding traffic scene. Note that we consider addressing the channel distortion effects, including dynamic distortion and multi-user interference, to improve the mean Intersection over Union (mIoU) of final fusion results performance
\item We propose a diffusion model (DM)-based channel enhancer (DMCE) to address the channel distortion by learning the distribution of data and received signals to suppress the noise in the CSI estimation. Based on the enhanced CSI estimation, the receiver can restore more accurate transmitted signals via channel equalization, which further improves the performance of semantic segmentation accuracy and mIoU of semantic image reconstruction.
\item We evaluated the performance of the proposed DMCE-based multi-user semantic communication scheme with a multi-source dataset in a traffic scenario. Simulation results show that the proposed DMCE is robust against dynamic channels, especially under low signal-to-noise ratio (SNR) conditions. Compared with the ablation experiment and benchmark schemes, the proposed scheme can improve the mIoU by 25.9\% and 39\% at 0 dB SNR, respectively.
\end{itemize}
\vspace{-0.1cm}

The rest of the paper is organized as follows. The system model is introduced in Section II. The design of the proposed DMCE-based multi-user semantic communication system is illustrated in Section III. Then, the simulation and analysis are presented in Section IV, and Section V concludes this paper.

\section{System model}\label{system model}

In this section, we introduce the components of the proposed DMCE-based multi-user semantic communication systems, as depicted in Fig. \ref{system structure}. We assume there are \(N\) user equipments (UEs) equipped with single-antenna, and each UE sends single-source image data, denoted by  \(\boldsymbol{m}_i, i=1,2,...,N\), to the centralized receiver. The centralized receiver is equipped with $M$ antennas. In the considered model, UEs send the semantic features of the multi-source images of the same traffic scene to a centralized receiver for transmitted semantic features recovery and fusion. Channel estimation and equalization are conducted at the receiver to compensate for the distortion of the wireless channel. Finally, the semantic features from different sources are fused at the receiver to obtain an aggregate semantic image of the traffic scene. The semantic image is the semantic segmentation of vehicles, pedestrians, etc., used to identify obstacles during vehicle driving to assist autonomous driving. Due to the inherent uncertainty of the wireless channel, the accuracy of  CSI estimated by the receiver directly determines the accuracy of the reconstructed semantic image.

\vspace{-0.1cm}

\subsection{Transmission Model}\label{Transmission}
As shown in Fig. \ref{system structure}, the semantic features contained in $\boldsymbol{m}_i$ is \(\boldsymbol{f}_i\in \mathbb{R} ^L,i=1,2,...,N\), where \(L\) is the length of the semantic features, and $\boldsymbol{f}_i$ is obtained by a semantic encoder. The JSCC encoder encodes the semantic features $\boldsymbol{f}_i$ to generate $\tilde{\boldsymbol{x}}_{\boldsymbol{i}}\in \mathbb{C} ^K,i=1,2,...,N$, and $K$ is the number of transmitted complex symbols. The semantic features $\boldsymbol{f}_i$ and the transmitted symbols $\tilde{\boldsymbol{x}}_{\boldsymbol{i}}$ can be respectively, given by
\vspace{-0.1cm}
\begin{equation}
\boldsymbol{f}_{\boldsymbol{i}}=\boldsymbol{F}_S\left( \boldsymbol{m}_i \right) ,
\label{eq}
\end{equation}
\begin{equation}
\tilde{\boldsymbol{x}}_{\boldsymbol{i}}=\boldsymbol{F}_C\left( \boldsymbol{f}_{\boldsymbol{i}} \right) ,
\label{eq}
\end{equation}

where $\boldsymbol{F}_S\left( \cdot \right), \boldsymbol{F}_C\left( \cdot \right) $  are the functions of the semantic encoder and the JSCC encoder, respectively. To satisfy the average power constraint $\frac{1}{K} \mathbb{E} _{\tilde{\boldsymbol{x}}}\left[ \left\| \tilde{\boldsymbol{x}} \right\| _{2}^{2} \right] \le P$, where $\left\| \cdot \right\|$ represents the L2 norm. The channel input symbols are derived by normalizing the coded symbols, which can be given by
\begin{equation}
\boldsymbol{x}_i=\sqrt{PK}\frac{\tilde{\boldsymbol{x}}_i}{\left\| \tilde{\boldsymbol{x}}_i \right\| _2}.
\label{eq}
\end{equation}

\begin{figure*}[t]
\centerline{\includegraphics[width = .73\textwidth]{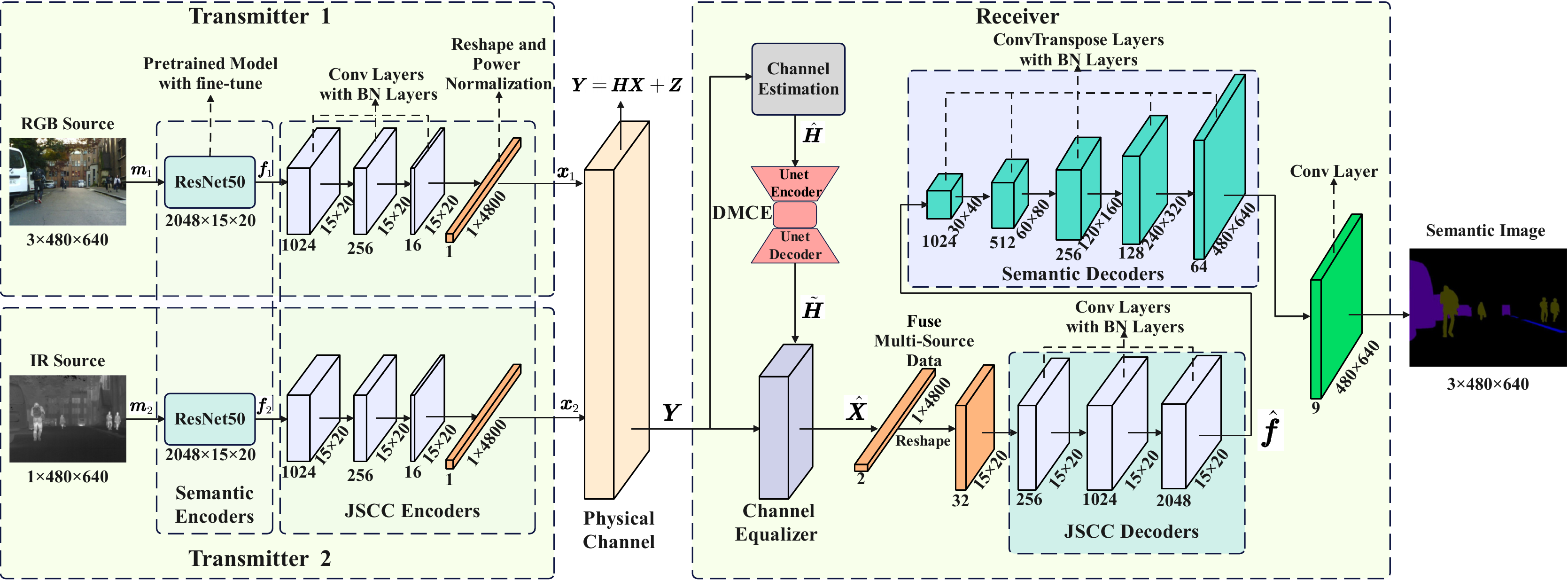}}
\caption{The network architecture for the proposed DMCE-based multi-user semantic communication systems.}
\label{network_architecture}
\vspace{-0.6cm}
\end{figure*}

Let $\boldsymbol{X}=\left[ \boldsymbol{x}_1,\boldsymbol{x}_2,...,\boldsymbol{x}_N \right] ^{\mathrm{T}}\in \mathbb{C} ^{N \times K}$ be the stacked matrix of all UEs, transmitted through a MIMO channel for space-division multiplexing, the output of the channel can be given by 
\begin{equation}
\boldsymbol{Y}=\boldsymbol{HX}+\boldsymbol{Z},
\label{eqY}
\end{equation}
where $\boldsymbol{Y}=\left[ \boldsymbol{y}_1,\boldsymbol{y}_2,...,\boldsymbol{y}_{M} \right] ^{\mathrm{T}}\in \mathbb{C} ^{M\times K}$. The CSI matrix is represented as $\boldsymbol{H}=\left[ \boldsymbol{h}_1,\boldsymbol{h}_2,...,\boldsymbol{h}_N \right] \in \mathbb{C} ^{M\times N}$, with $\boldsymbol{h}_i=\left[ h_{1i},...,h_{ji},...,h_{Mi} \right] ^{\mathrm{T}}$ being the channel vector and $h_{ji}$ being the channel gain between the $i$-th transmitter and the $j$-th receiver antenna. In (\ref{eqY}),   $\boldsymbol{Z}\in \mathbb{C} ^{M\times K}$ represents the additive noise terms that follow the independent identically distributed (i.i.d.) circular symmetric complex Gaussian (CSCG) distributions with zero mean and variance $\sigma^2$, which is denoted by \(z_{i, j}\sim \mathcal{C} \mathcal{N} \left( 0,\sigma ^2 \right)\).

After the channel equalization, the received signal is expressed as 
\vspace{-0.2cm}
\begin{equation}
\hat{\boldsymbol{X}}=\left[ \left( \tilde{\boldsymbol{H}} \right) ^{\mathrm{H}}\tilde{\boldsymbol{H}} \right] ^{-1}\left( \tilde{\boldsymbol{H}} \right) ^{\mathrm{H}}\boldsymbol{Y}=\check{\boldsymbol{X}}+\hat{\boldsymbol{Z}},
\label{eqXhat}
\end{equation}
where $\check{\boldsymbol{X}}=\left[ \left( \tilde{\boldsymbol{H}} \right) ^{\mathrm{H}}\tilde{\boldsymbol{H}} \right] ^{-1}\left( \tilde{\boldsymbol{H}} \right) \boldsymbol{X}$ is the restored transmitted symbols, and the residual noise is $\hat{\boldsymbol{Z}}=\left[ \left( \tilde{\boldsymbol{H}} \right) ^{\mathrm{H}}\tilde{\boldsymbol{H}} \right] ^{-1}\left( \tilde{\boldsymbol{H}} \right) \boldsymbol{Z}$. Moreover, the enhanced CSI estimation, $\tilde{\boldsymbol{H}}$, is enhanced by DMCE based on the initial CSI estimation, $\hat{\boldsymbol{H}}$, as shown in Fig. \ref{system structure}. The enhanced CSI estimation can be given by
\vspace{-0.1cm}
\begin{equation}
\tilde{\boldsymbol{H}}=\boldsymbol{F}_D\left( \hat{\boldsymbol{H}} \right) ,
\label{eq_H}
\end{equation}
where $\boldsymbol{F}_D\left( \cdot \right) $  is the function of DMCE and will be elaborated in Section \ref{Sec3}. Then, $\hat{\boldsymbol{X}}$ are fed into the JSCC decoder to recover the transmitted semantic features, and we obtain$\hat{\boldsymbol{f}}_i\in \mathbb{R} ^L,i=1,2,...,N$, which can be expressed as
\vspace{-0.1cm}
\begin{equation}
\hat{\boldsymbol{f}}=\boldsymbol{F}_{C}^{I}\left( \hat{\boldsymbol{X}} \right) ,
\label{eq}
\end{equation}
where $\hat{\boldsymbol{f}}$ is semantic features, and $\boldsymbol{F}_{C}^{I}\left( \cdot \right) $ is the function of the JSCC decoder. Then the multi-source semantic features $\hat{\boldsymbol{f}}=\left[ \hat{\boldsymbol{f}}_{1}^{\mathrm{T}},...,\hat{\boldsymbol{f}}_{N}^{\mathrm{T}} \right] $ is fed into the semantic decoder to  obtain the ultimate fused result, which can be given by 
\vspace{-0.1cm}
\begin{equation}
\hat{\boldsymbol{\Gamma}}=\boldsymbol{F}_{S}^{I}\left( \hat{\boldsymbol{f}} \right) ,
\label{eq}
\end{equation}
where $\boldsymbol{F}_{S}^{I}\left( \cdot \right) $  is the function of the semantic decoder, and $\hat{\boldsymbol{\Gamma}}$  is the inferred semantic segmentation images.

\subsection{Channel Model} \label{Channel Model}
In this paper, we adopt the ray-tracing channel model, which can be given by \cite{chen2021}
\begin{equation}
\boldsymbol{H}(f) = \sum_{\ell=1}^{L} \alpha_{\ell}e^{j2\pi(-\tau_{\ell}f)}\mathbf{a}_{rx}\left(\phi^{(\ell)}_{rx}, \theta^{(\ell)}_{rx}\right)\mathbf{a}^{H}_{tx}\left(\phi^{(\ell)}_{tx}, \theta^{(\ell)}_{tx}\right),
\label{eq}
\end{equation}
where $\boldsymbol{H}(f)$ represents the frequency-domain channel response, $\alpha_{\ell}$ is the channel gain of path $\ell$, $\tau_{\ell}$ is the path delays, $f$ is the frequency, $\mathbf{a}_{rx}$ and $\mathbf{a}^{H}_{tx}$ are the receiver and transmitter steering vectors, respectively, and $\phi$ and $\theta$ are the azimuth and elevation angles, respectively. After the transmitters and the centralized receiver finish the beam alignment procedures, the frequency-domain channel response  can be transformed into 
\vspace{-0.2cm}
\begin{equation}
\boldsymbol{H}(f) = \sum_{\ell=1}^{L} \alpha_{\ell}e^{j2\pi(-\tau_{\ell}f)}G_{rx}G_{tx} ,
\label{eq}
\end{equation}
where $G_{rx}$ and $G_{tx}$ represent the receiving and transmitting beamforming gains, respectively. 

At the receiver, the CSI matrix estimated via pilot signals can be given by
\begin{equation}
\hat{\boldsymbol{H}} = \boldsymbol{H} + \boldsymbol{Z}_H
 ,
\label{eqHhat}
\end{equation}
where $\boldsymbol{Z}_H \sim \mathcal{C} \mathcal{N} \left( 0, \sigma_H^2\boldsymbol{I} \right) $  is the complex noise terms following i.i.d. CSCG distribution, and $\sigma_H^2$ is the power of the noise. More accurate CSI estimation can recover the signal $\hat{\boldsymbol{X}}$ with higher accuracy as shown in (\ref{eqXhat}). Therefore, we can enhance $\hat{\boldsymbol{H}}$ using the DMCE mentioned in (\ref{eq_H}) and obtain $\tilde{\boldsymbol{H}}$, where the noise in $\hat{\boldsymbol{H}}$ can be greatly suppressed.

\section{DMCE-based multi-user semantic communication systems} \label{Sec3}
In this section, we present the multi-user semantic communication system mechanism that uses DMCE to enhance the physical channel equalization as shown in Fig. \ref{network_architecture}. In the system, Multiple users use the semantic encoder to compress the multi-source images and generate semantic latent vectors to transmit through wireless channels with distortion. The receiver receives the distorted signals containing the latent vectors and applies a channel equalizer to eliminate multi-user interference and intersymbol interference as shown in (\ref{eqXhat}). Here, the DMCE is used to improve the CSI estimation and further enhance the channel equalization, finally improving the decoding accuracy.

\subsection{Semantic Encoder at the Transmitter}\label{AA}
As illustrated in Fig. \ref{network_architecture}, the semantic features transmitted by multiple UEs are fused naturally by the JSCC decoder and consist of data from various sources $\boldsymbol{m}_i, i=1,2,...,N$ , $m_i\in \mathbb{R} ^{B\times C\times 480\times 640}$. Here, $B$  and $C$ represent the batch size and channel number of the input image, respectively, and the height and width of the images are 480 and 640, respectively. The semantic encoder utilizes the ResNet-50 framework, which has been pretrained based on the ImageNet dataset \cite{Deng2009}. The semantic features, $\boldsymbol{f}_i\in \mathbb{R} ^{B\times 2048\times 15\times 20}$,  are mapped into the symbols, and we obtain  $\tilde{\boldsymbol{x}}_{\boldsymbol{i}}\in \mathbb{C} ^{B\times 1\times 1\times 4800}$, via a JSCC encoder. This JSCC encoder is composed of three CNN blocks, each containing a convolutional layer, a batch normalization layer, and a LeakyRelu layer. To meet power constraints, a normalization layer is utilized to normalize $\tilde{x}_i$ into $x_i$.  Subsequently, $\boldsymbol{x}_{\boldsymbol{i}}$ is transmitted through the wireless channel to the centralized receiver.

\subsection{DMCE-enabled Semantic Decoder at the Receiver}
The receiver comprises the channel equalizer, DMCE, JSCC decoder, and semantic decoder modules. Specifically, the receiver initially receives the raw signals $\boldsymbol{Y}\in \mathbb{C} ^{B\times 1\times 2\times 4800}$. The DMCE module is composed of a well-trained conditional DM.  The initial CSI estimation, $\hat{\boldsymbol{H}}$, as shown in (\ref{eqHhat}), serves as the conditional input to the DMCE module, yielding the enhanced CSI estimation $\tilde{\boldsymbol{H}}$. The enhanced CSI estimation  $\tilde{\boldsymbol{H}}$ enables the channel equalizer to obtain more accurate $\tilde{\boldsymbol{X}}$ from $\boldsymbol{Y}$. The JSCC decoder consists of three CNN blocks. The JSCC decoder processes $\tilde{\boldsymbol{X}}$ to reconstruct the semantic features of all sources, i.e., $\hat{\boldsymbol{f}}\in \mathbb{R} ^{B\times 2048\times 15\times 20}$. Since multi-source images transmitted by multiple users represent the same scene as shown in Fig. \ref{network_architecture}, there is no need to restore the semantic features for each of the $N$ transmitters separately. Instead, the semantic features transmitted by multiple UEs are concated and fused naturally by the JSCC decoder, and the semantic features  $\hat{\boldsymbol{f}}$ are directly fed into the semantic decoder to obtain the final semantic image. The semantic decoder comprises five CNN blocks and a final output layer. Each of the first five CNN blocks consists of a convolutional layer with a stride of 2, a batch normalization layer, and a LeakyRelu layer. The output layer is implemented by a CNN layer, producing the final semantic segmentation images $\hat{\Gamma}$.
\vspace{-0.2cm}
\subsection{DMCE Description}
The DM reconstructs the desired data's distribution by iteratively learning the distribution of noise and removing the noise from the desired signals. Specifically, in the forward diffusion process of DM, DM gradually adds Gaussian noise to the training data until the perturbed signals approach pure noise. Then, in the reverse sampling process, DM learns to recover the true data signals from the perturbed signals. Here, the receiver needs to enhance the initial CSI estimation. Therefore, we need to employ a conditional DM. In the forward diffusion process, we regard the perfect CSI matrix $\boldsymbol{H}$ as the original data $\mathrm{x}_0$. Then, the original data distribution can be denoted as $\mathrm{x}_0 \sim q\left( \boldsymbol{H} \right) $ , and the forward diffusion process can generate the  $t$-th sample,   $\mathrm{x}_t$, by sampling a Gaussian vector $\epsilon \sim \mathcal{N} \left( 0,\mathbf{I} \right)$  as follows:
\begin{equation}
\mathrm{x}_{\mathrm{t}}=\sqrt{\bar{\alpha}_{\mathrm{t}}}\mathrm{x}_0+\sqrt{1-\bar{\alpha}_{\mathrm{t}}}\epsilon,
\label{sum-rate}
\end{equation}
where $\bar{\alpha}_t=\prod_{i=1}^t{\left( 1-\beta _i \right)}$ , and $\beta _i=1-\alpha_i, i=1,…,t$  are predefined variance schedules. The forward diffusion process of DM can be represented by the following distribution
\begin{equation}
q(\mathrm{x}_t|\mathrm{x}_0) \sim \mathcal{N} (\mathrm{x}_t;\sqrt{\bar{\alpha}_t}\mathrm{x}_0,(1-\alpha_t)\boldsymbol{I}).
\label{beam_search}
\end{equation}

The reverse sampling process intends to sample the original data distribution from  CSI estimation $\hat{\boldsymbol{H}}$, which means $\mathrm{x}_t=\hat{\boldsymbol{H}}$. The posterior probability chain can be represented as
\begin{equation}
p_{\theta}\left( \mathrm{x}_{t-1}\left| \mathrm{x}_t \right. \right) \sim \mathcal{N} \left( \mathrm{x}_{t-1};\mu \left( \mathrm{x}_t \right) , \sigma _{\theta}\left( \mathrm{x}_t \right) \boldsymbol{I} \right),
\label{reverse_process}
\end{equation}
where $\theta$  represents the training parameters of the DM, $\mu \left( \mathrm{x}_t \right) =\frac{1}{\alpha _t}\left( \mathrm{x}_t-\frac{\beta _{t}^{2}}{\bar{\beta}_t}\epsilon _{\theta}\left(\mathrm{x}_t,t \right) \right) 
$ , and $\sigma _{\theta}\left( \mathrm{x}_t \right)$  is the variance obtained by DM, $\epsilon _{\theta}\left(\mathrm{x}_t, t \right)$  is the noise predicted by DM. The DM can be trained by minimizing the loss between the predicted noise and the true noise, and the loss function is given by
\begin{equation}
\mathbb{E} _{\mathrm{x}_0,\varepsilon \sim \mathcal{N} \left( 0,\boldsymbol{I} \right)}\left[ \left\| \varepsilon -\epsilon _{\theta}\left( \bar{\alpha}_t\mathrm{x}_0+\bar{\beta}_t\varepsilon ,t \right) \right\| ^2_2 \right] .
\label{sum-rate}
\end{equation}

\subsection{Training Strategy}
The proposed system is trained using a three-stage strategy. In the first stage, we train the semantic encoder and decoder, JSCC encoder and decoder that ignore the noise $\boldsymbol{Z}$. The ResNet-50, pre-trained on the ImageNet dataset, is fine-tuned to align with the tasks. Specifically, the IR data have one channel. As a result, the number of input channels of the first convolutional layer of the pre-trained ResNet-50 model used for extracting infrared is modified to 1. Additionally, the final linear layer of each pre-trained ResNet-50 model is discarded to retain more extracted features. The semantic encoders and decoders of RGB and IR sources are trained by the same cross-entropy loss function, which is represented as
\begin{equation}
\mathcal{L} _{SC}=\sum_{g=1}^{G}{\Gamma _{g}\log \left( \hat{\Gamma}_{g} \right)}+\left( 1-\Gamma _{g} \right) \log \left( 1-\hat{\Gamma}_{g} \right) ,
\label{eqCE}
\end{equation}
where $G$ is the number of object categories in the dataset. $\Gamma _{g}$ is the actual probability of category $g$ , and $\hat{\Gamma}_{g}$  is the model's predicted probability for category $g$.

In the second stage, we introduce the conditional DM into the network. First, the model parameters trained in the first stage are loaded, and these parameters are no longer updated. The perfect CSI estimation, $\boldsymbol{H}$, is assumed to follow the distribution $\mathrm{x}_0\sim q\left( \boldsymbol{H} \right)$  that needs to be learned. Initial CSI estimation, $\hat{\boldsymbol{H}}$, is used as the input for the conditional DM. Then, the conditional DM can be trained by minimizing the Mean Squared Error (MSE) between the actual noise and the noise predicted by the model, which is represented as 
\begin{equation}
\mathcal{L} _{DM}= \\
\mathbb{E} _{\mathrm{x}_0,\varepsilon}\left[ \left\| \varepsilon -\epsilon _{\theta}\left( \bar{\alpha}_t\mathrm{x}_0+\bar{\beta}_t\varepsilon , \thinspace t \right) \right\| _{2}^{2} \right],
\label{eq}
\end{equation}
where $\varepsilon$ is the actual noise, and $\epsilon _{\theta}\left( \bar{\alpha}_t\mathrm{x}_0+\bar{\beta}_t\varepsilon ,\thinspace t \right) $ is the noise predicted by the model. After the model predicts noise, we can denoise $\hat{\boldsymbol{H}}$ based on (\ref{reverse_process}).

In the third stage, the entire network is jointly trained again. During this phase, the parameters of the semantic encoder, JSCC encoder, and conditional DM are frozen, while the parameters of the JSCC decoder and semantic decoder are updated to obtain the optimal solution for the entire system. The loss function $\mathcal{L} _{SC}$ in the first stage is used to update their parameters. The overall training process of the system is summarized in \textbf{Algorithm} 1.

\begin{algorithm}[t]
	\caption{The joint training algorithm}
	\label{alg:joint training}
	\begin{algorithmic}[1]
		\Require The multi-source dataset $\boldsymbol{M}$;
		\Ensure The well-trained DMCE for multi-user semantic communication systems;
		\While{loss is not converged}
		\State Random sample $\boldsymbol{m}$ from the dataset $\boldsymbol{M}$. 
            \State Perform forward propagation without conditional DM.
            \State Compute $\mathcal{L} _{SC}$ and update $F_S\left( \cdot \right) $, $ F_C\left( \cdot \right) $, $ F_S^{I}\left( \cdot \right) $, $ F_C^{I}\left( \cdot \right) $.
        \EndWhile
        \While{loss is not converged}
		  \State Randomly generate $\boldsymbol{H}$.
		  \State Perform forward propagation with conditional DM.
            \State Compute $\mathcal{L} _{DM}$ and update $F_D\left( \cdot \right) $.
		\EndWhile
        \While{loss is not converged}
		  \State Random sample $\boldsymbol{m}$ from the dataset $\boldsymbol{M}$.
		  \State Perform forward propagation with conditional DM.
            \State Compute $\mathcal{L} _{SC}$ and update $F_S^{I}\left( \cdot \right) , F_C^{I}\left( \cdot \right) $.
		\EndWhile
	\end{algorithmic}
\end{algorithm}
\vspace{-0.1cm}
\section{Performance evaluation}
In this section, we present the visualized results, the channel enhancement via DMCE, and the mIoU of the inferred image results to show the feasibility and effectiveness of our proposed DMCE-based multi-user semantic communication systems.

\subsection{Experimental Settings}\label{Experimental Settings}
\paragraph{Datasets} The multi-source image dataset used in the experiments is referred to \cite{Ha2017}, comprising 1569 pairs of RGB-IR urban traffic scene images. The dataset performs semantic segmentation on nine classes of objects, including unlabelled pixels and eight common objects in traffic environments (car, person, bike, curve, etc.). Unlabelled pixels account for 92.138\% of the dataset, while the remaining 7.862 \% pixels correspond to the other eight objects. Both RGB and IR images have a resolution of 480×640 pixels. RGB images contain three color channels, while IR images have one grayscale channel.

\paragraph{Training parameters} During the training process of \textbf{Algorithm} 1, the noise addition step count $T$ was set to 1000, with $\beta_i$ linearly increasing from $\beta_1=0.0001$ to $\beta_T=0.02$. The noise prediction model used in DMCE is an attention-Unet. In the sampling process of DMCE, the sampling step size was set to $t_s =  arg\underset{t}{\min}\sigma^2-\frac{1-\bar{\alpha}_t}{\bar{\alpha}_t}$ \cite{Wu2023}, with the loss function being the MSE between the predicted noise and the actual noise. For the first and third training stages in \textbf{Algorithm} \ref{alg:joint training}, a Stochastic Gradient Descent (SGD) optimizer with an initial learning rate of 0.01 and a decay rate of 0.95 is utilized. Each training stage was trained for 100 epochs, and the model with the best performance on the validation set was recorded and used for the test. The loss function of stage1 and stage3 in \textbf{Algorithm} \ref{alg:joint training} is shown in (\ref{eqCE}).

\subsection{System Metric}\label{Problem Description}
In this paper, we conduct tests under various values of SNR to evaluate the system's performance in recovering semantic information and demonstrate the robustness of our proposed solution over various SNRs. The multi-source image information is fused at the receiver to produce semantic images. We utilize the mIoU as the system performance evaluation metric, which can be given by
\begin{equation}
\mathrm{mIoU}=\frac{1}{v+1}\sum_{i=0}^v{\frac{p_{ii}}{\sum{_{i=0}^{v}p_{ij}+}\sum{_{i=0}^{v}p_{ji}-p_{ii}}}},
\label{eq}
\end{equation}
where $v$ is the total number of classifications, $p_{ij}$ represents the false prediction of class $i$ as class  $j$, i.e., false negative (FN), $p_{ji}$ represents the false prediction of class $j$ as class  $i$, i.e., False Positive (FP), and $p_{ii}$ represents the correct prediction of class $i$ as class  $i$, i.e., True Positive (TP).

\subsection{Simulation Results}
\begin{figure}[t]
% \vspace{-0.25cm}
\centerline{\includegraphics[width = .38\textwidth]{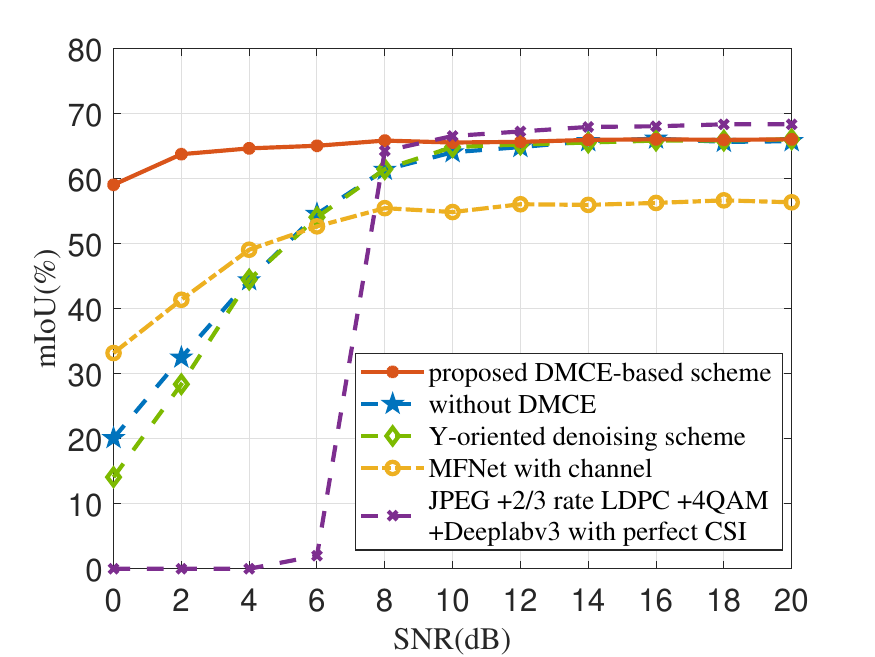}}
\setlength{\abovecaptionskip}{-0.15cm}
\caption{mIoU performance at various SNRs.}
\label{mIoU}
\vspace{-0.4cm}
\end{figure}

\begin{figure}
% \vspace{-0.1cm}
\centerline{\includegraphics[width = .38\textwidth]{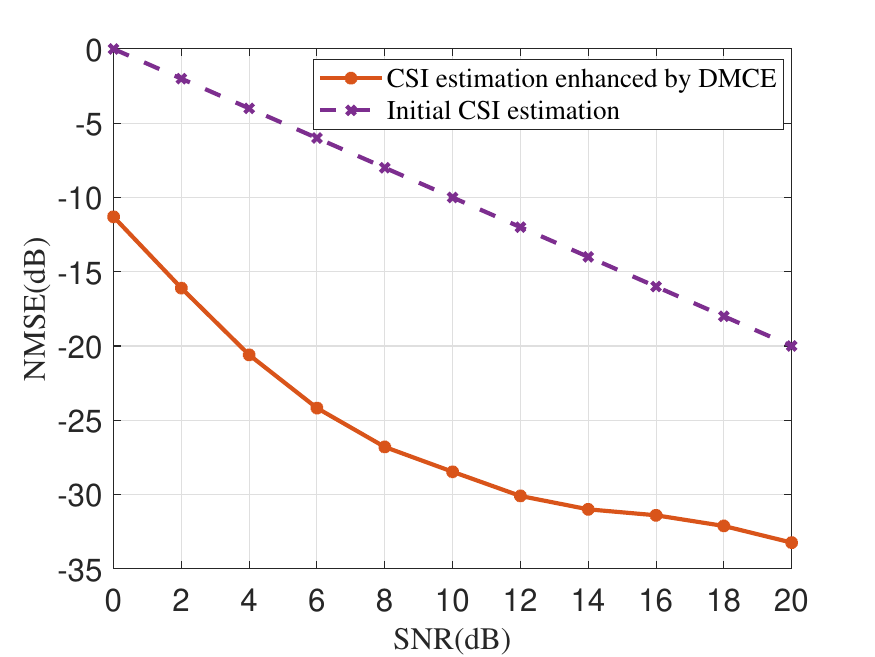}}
\setlength{\abovecaptionskip}{-0.15cm}
\caption{DMCE can enhance the initial CSI estimation NMSE value at various SNRs.}
\label{Enhance_H}
\vspace{-0.3cm}
\end{figure}

\begin{figure}[t]
\centerline{\includegraphics[width = .4\textwidth]{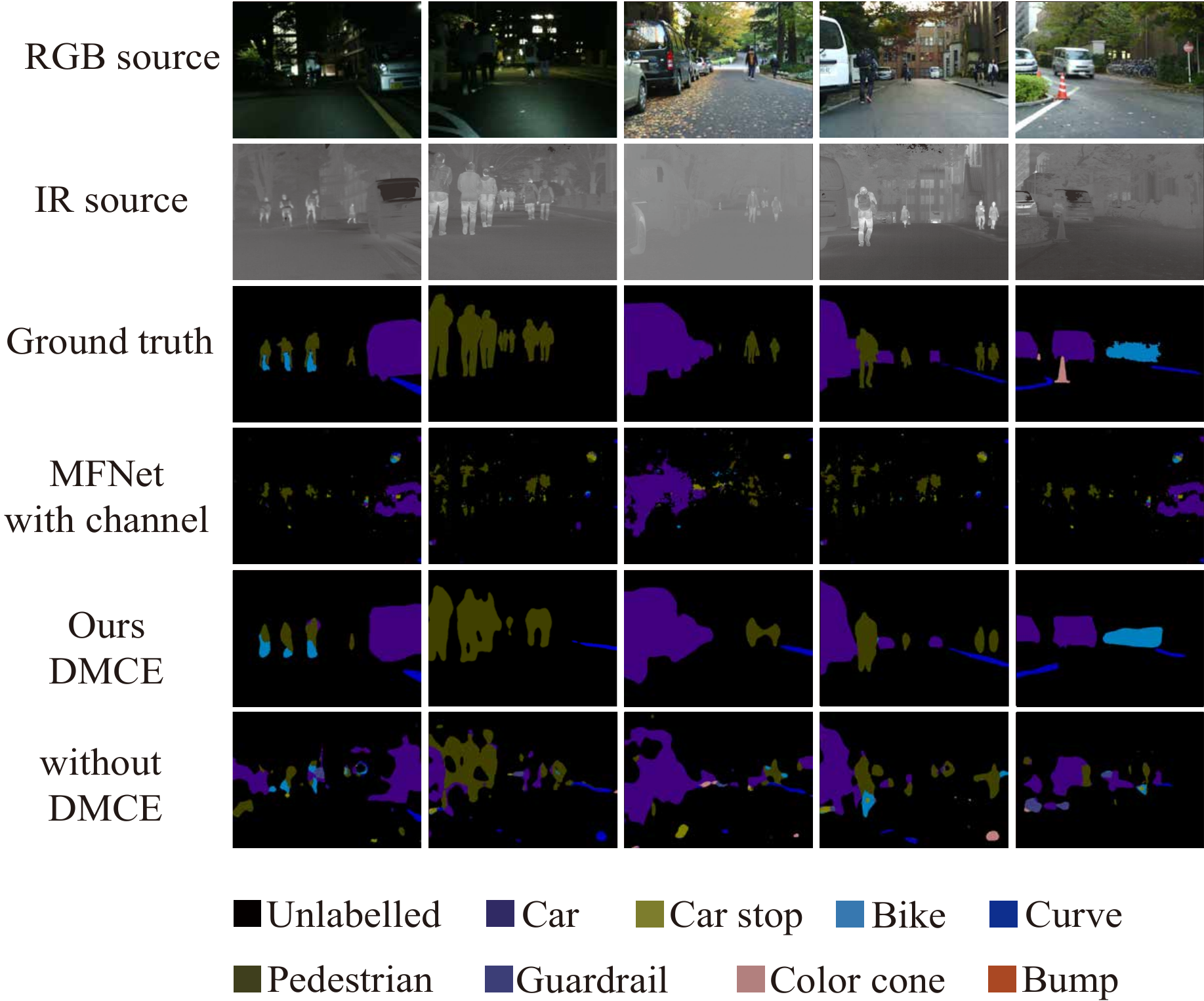}}
\caption{The inferred results of the DMCE-based scheme and comparison schemes under SNR of 4 dB. All the results are inferred on test sets. }
\label{visualize}
\vspace{-0.4cm}
\end{figure}

In Fig. \ref{mIoU}, we present the mIoU performance of our proposed DMCE-based scheme and the comparison schemes under various SNRs. For comparison, we apply the ray-tracing channel model described in Section \ref{Channel Model} to the MFNet \cite{Ha2017} scheme, using it as a benchmark for performance comparison. From Fig. \ref{mIoU} we can see that DMCE-based scheme can maintain a high mIoU performance of around 65\% under various SNRs, while the baselines have lower mIoU when the SNR is lower than 8 dB. This is because the proposed scheme greatly enhances the CSI estimation at low SNR regimes, further improving semantic information recovery. We also tested the effect of using DM to denoise only $\boldsymbol{HX}$ in the Y-oriented experiments. As shown in Fig \ref{mIoU}, the Y-oriented scheme is not as effective as the DMCE scheme. This is because the BatchNormalization layers in the networks cause the distribution of $\boldsymbol{HX}$ and the noise to both be normal distributions, which does not allow for effective data denoising. Fig. \ref{mIoU} also shows that, when the SNR is higher than 6 dB, the mIoU of the traditional digital coding scheme, which uses JPEG+2/3 rate LDPC scheme with perfect CSI estimation, experiences a sudden increase and is slightly higher than the proposed scheme. The sudden improvement is due to the cliff effect \cite{Pan2023}. At high SNR, the traditional scheme outperforms the proposed DMCE scheme in mIoU because of JPEG's reliable data bit transmission, with a cost of 96 times of transmission amount than DMCE-based scheme. Fig. \ref{mIoU} demonstrates that the proposed DMCE-based scheme can enhance the accuracy of CSI estimation, especially under low SNR conditions.

Fig. \ref{Enhance_H} shows the performance of CSI estimation enhancement of DMCE under various SNRs. It can be seen that DMCE can significantly decrease the normalized MSE (NMSE) of the CSI estimation compared with the initial one under various SNRs. Specifically, DMCE reduces the CSI estimation NMSE by an average of 14 dB, which further explains the improvement of mIoU performance and visual results in Figs. \ref{mIoU} and \ref{visualize}, respectively. 

Fig. \ref{visualize} represents the visual results of the proposed DMCE-based scheme and other schemes with the test set under SNR of 4 dB. Since the results of JPEG+2/3 rate LDPC scheme are total noise, we omit the visual results of this scheme. The initial CSI estimation without enhancement is applied in the ablation experiment without DMCE. We chose 2 images taken at night and 3 images taken during the day for evaluation. It can be seen that the proposed DMCE-based scheme ensures the best performance of multi-source image fusion for semantic segmentation at 4 dB SNR since the DMCE can enhance the CSI estimation and further improve the recovery of the semantic information in equalization processing. However, the comparison schemes can barely recognize the results as the semantic features are distorted completely by poor channel conditions.

\vspace{-0.2cm}
\section{Conclusion}
In this paper, we proposed a DMCE-based multi-user semantic communication system that exploits DM to enhance the CSI estimation for improving the accuracy of the restoration of the semantic features via channel equalization at the receiver. Extensive evaluation and simulation are conducted based on a real-world image dataset. The simulation results show that the DMCE-based scheme outperforms the benchmark schemes by more than 25\% in mIoU performance,  which shows the feasibility and effectiveness of the proposed system in recovering semantic image tasks. The proposed scheme holds promise for application in various automation applications, including autonomous vehicle navigation.

%\begin{thebibliography}{1}
% \def\baselinestretch{0.6}
% \bibliographystyle{IEEEtran}
% \bibliography{IEEEabrv,refs}
%\end{thebibliography}

\def\baselinestretch{0.8}
\bibliographystyle{IEEEtran}
\bibliography{IEEEabrv,refs}

\end{document}